\documentclass[final,3p,times,authoryear]{elsarticle}
\usepackage{amsmath,amssymb,amsfonts}
\usepackage{amsthm}
\usepackage{booktabs}
\usepackage{graphicx}
\usepackage{microtype}
\usepackage{algorithm}
\usepackage{algpseudocode}
\usepackage{url}
\usepackage{hyperref}

\hypersetup{
  colorlinks=true,
  linkcolor=blue,
  citecolor=blue,
  urlcolor=blue
}

\makeatletter
\def\ps@pprintTitle{%
  \let\@oddhead\@empty
  \let\@evenhead\@empty
  \let\@oddfoot\@empty
  \let\@evenfoot\@empty
}
\makeatother

\newtheorem{theorem}{Theorem}
\newtheorem{proposition}{Proposition}

\theoremstyle{definition}

\newtheorem{remark}{Remark}

\begin{document}

\begin{frontmatter}

  \title{Revisiting Generalized Median Estimators:\\
  Exact Bitwise Computation and Gray Code Acceleration}

  \author{Robin Nunkesser}
  \ead{robin.nunkesser@hshl.de}
  \address{Department of Computer Science, Hamm-Lippstadt University
  of Applied Sciences, Marker Allee 76--78, 59063 Hamm, Germany}

  \begin{abstract}
    Generalized median estimators evaluate the median of a symmetric
    kernel over all $M = \binom{n}{k}$ subsets of size $k$. Exhaustive
    evaluation was considered impractical even for $n = 29, k = 5$
    ($M = 118,755$), prompting stochastic subsampling (Wilde and
    Grimshaw, 2013). We show that this barrier is computational, not
    combinatorial: storing all subsets requires $\Theta(kM)$ space,
    whereas stateful enumeration uses $\mathcal{O}(k)$ auxiliary
    state and retains only $M$ kernel values. We combine two
    exhaustive traversals with three-way Quickselect and a
    deterministic median-of-medians fallback, giving worst-case
    $\mathcal{O}(M)$ selection. Gosper's successor takes
    $\mathcal{O}(k)$ kernel work per subset for $n < w$; a Revolving
    Door Gray code changes one element out and one in, enabling
    $\mathcal{O}(1)$ kernel updates between recalibrations and
    $\mathcal{O}(M + kM/P)$ total time. We derive floating-point
    error bounds for the location ($h_1$) and mean-centered sample
    variance ($h_2$) kernels and their resulting median. For
    $n = 29, k = 5$, our ANSI C implementation evaluates all
    $118,755$ subsets in $0.74$ ms and $0.71$ ms, respectively. A
    direct C sweep places the crossover with partial lexicographic
    unranking at approximately $18,000$--$23,000$ subsets
    ($15\%$--$19\%$ of $M$), below the $32,160$--$45,420$ subsets
    required by Wilde and Grimshaw's sequential approximation at
    $\epsilon = 0.01$.
  \end{abstract}


\end{frontmatter}

\section{Introduction}
\label{sec:intro}

Robust location and scale estimation frequently exhibits an inherent
combinatorial structure. Whereas classical $L_2$-based moment
estimators such as the sample mean and sample variance can be
evaluated in linear time $\mathcal{O}(n)$ by single-pass
accumulation, high-breakdown estimators typically inspect
multi-element subsets of observations. Prominent examples include the
Hodges-Lehmann location estimator \citep{HodgesLehmann1963}, Least
Median of Squares (LMS) and Least Trimmed Squares (LTS) regression
\citep{Rousseeuw1984,Stromberg1993}, repeated median filters
\citep{BernholtFried2003,BernholtFriedGatherWegener2006}, and the
pairwise scale estimators $S_n$ and $Q_n$
\citep{CrouxRousseeuw1992,RousseeuwCroux1993,SchettlingerNunkesserFried2008,NunkesserFriedGather2009}.

Generalized median (GM) estimators, introduced by
\citet{Serfling1984}, \citet{BrazauskasSerfling2000}, and
\citet{Serfling2002} and investigated computationally by
\citet{WildeGrimshaw2013}, define a general class of robust
$U$-quantile statistics that evaluate the median of a symmetric
$k$-argument kernel $h(x_{i_1}, \dots, x_{i_k})$ over all $M =
\binom{n}{k}$ subsets of size $k$ drawn from a univariate sample $X =
(x_0, x_1, \dots, x_{n-1}) \in \mathbb{R}^n$. Letting
$\mathcal{S}_{n,k} = \{ S \subseteq \{0, \dots, n-1\} : |S| = k \}$,
the canonical kernels proposed by \citet{Serfling2002} for location
and scale (variance) are:
\begin{equation}
  \label{eq:kernels}
  h_1(x_{i_1}, \dots, x_{i_k}) = \frac{1}{k} \sum_{r=1}^k x_{i_r}, \qquad
  h_2(x_{i_1}, \dots, x_{i_k}) = \frac{1}{k - 1} \sum_{r=1}^k
  (x_{i_r} - \bar{x}_S)^2 \quad (k > 1).
\end{equation}
For a general symmetric kernel $h \in \{h_1, h_2\}$, the generalized
median estimator $\text{GM}_{n,k}^{(h)}(X)$ is given by:
\begin{equation}
  \label{eq:gm}
  \text{GM}_{n,k}^{(h)}(X) = \operatorname{median}_{S \in
  \mathcal{S}_{n,k}} h(S),
\end{equation}
where we write $\text{GM}_{n,k}(X) \equiv \text{GM}_{n,k}^{(h_1)}(X)$
when referring to the location estimator. Specifically, ordering the
$M = \binom{n}{k}$ subset kernel evaluations as $\mu_{(0)} \le
\mu_{(1)} \le \dots \le \mu_{(M-1)}$, we evaluate the median order
statistic $\mu_{(\lfloor M/2 \rfloor)}$. Whenever $M$ is odd
(including the canonical benchmark $M = \binom{29}{5} = 118,755$),
$\mu_{(\lfloor M/2 \rfloor)}$ is the unique central median
$Q_h(0.5)$; for even $M$, we adopt the upper median $\mu_{(M/2)}$ (if
  the conventional midpoint $(\mu_{(M/2-1)} + \mu_{(M/2)})/2$ is
  desired, $\mu_{(M/2-1)}$ equals the maximum of the lower partition
  after selecting $\mu_{(M/2)}$ and can be obtained in a single
$\mathcal{O}(M)$ scan). Throughout this paper, we use the term
\emph{exact} in the sense of \citet{WildeGrimshaw2013} to mean
\emph{combinatorially exhaustive evaluation of all $M = \binom{n}{k}$
subsets} in IEEE-754 floating-point arithmetic, as opposed to
stochastic subsampling of $N^* < M$ subsets. As noted by
\citet{Serfling2002} and \citet[p.~308]{WildeGrimshaw2013}, the
choice of $k$ continuously tunes the trade-off between robustness and
efficiency: $k = 1$ yields the upper sample median, $k = 2$ for
$h_1(\cdot)$ yields a distinct-pair variant of the Hodges-Lehmann
estimator (the classical Walsh-average definition also includes pairs
$i=j$), and $k = n$ recovers the sample mean $\bar{X}$ (and sample
variance $s^2$ for $h_2$). In particular, \citet{Serfling2002}
recommended $k = 5$ as offering an overall balance of breakdown
point, gross error sensitivity, and asymptotic relative efficiency.
Whereas the pairwise case $k = 2$ (Hodges-Lehmann and $Q_n$) admits
implicit $\mathcal{O}(n \log n)$ selection in sorted $n \times n$
matrices \citep{CrouxRousseeuw1992} and high-dimensional regression
variants with $k = p + 1$ are NP-hard when the dimension $p$ grows
with the input \citep{Bernholt2006}, univariate
$\text{GM}_{n,k}^{(h)}$ for fixed $k \ge 3$ involves selecting among
$M = \binom{n}{k} = \mathcal{O}(n^k)$ kernel values (a $k$-SUM
selection problem for $h_1$), making ultra-fast enumeration of
$\mathcal{S}_{n,k}$ the central algorithmic challenge.

\subsection{The Dogma of Computational Infeasibility in
\texorpdfstring{\citet{WildeGrimshaw2013}}{Wilde and Grimshaw (2013)}}

Despite the appealing statistical properties of $\text{GM}_{n,k}$ for
$k = 5$, its practical adoption has been curtailed by the perception
that evaluating all $M = \binom{n}{k}$ subsets is intractable. In
their study on computing generalized median estimators,
\citet[p.~308]{WildeGrimshaw2013} motivated stochastic subsampling as follows:
\begin{quote}
  ``While the ability to tune the GM estimator makes it a valuable
  method, evaluation of the kernel at $\binom{n}{k}$ possible subsets
  quickly becomes computationally infeasible even for modest $n$. One
  intuitive approach is to select subsets based on simple random
  sampling. That is, for $N^* < \binom{n}{k}$, evaluate the kernel at
  $N^*$ randomly sampled subsets and approximate the GM estimate from
  the median of the sampled subsets.'' \citep[p.~308]{WildeGrimshaw2013}
\end{quote}
Throughout their paper, \citet[pp.~308--316]{WildeGrimshaw2013}
centered their analysis on a benchmark dataset of $n = 29$
observations with $k = 5$, yielding $\binom{29}{5} = 118,755$ subsets
(with exact values $\hat{\mu} = 5.330$ for $h_1$ and $\hat{\sigma}^2
= 0.6933$ for $h_2$). After demonstrating that simple random sampling
\emph{with replacement} fails---noting on p.~311 that even at $N^* =
200,000$ (nearly double $\binom{29}{5} = 118,755$), approximate
estimates still exhibit non-negligible variation from $\hat{\mu} =
5.330$---they turned to sampling \emph{without replacement}.
Crucially, when discussing candidate algorithms for subset generation
without replacement, \citet[p.~312]{WildeGrimshaw2013} explicitly
considered and rejected incremental subset enumeration:
\begin{quote}
  ``Another greedy approach is to generate and store the sample frame
  containing all possible subsets. This approach can be made more
  efficient by sequentially generating the next subset by performing
  a small change to the previous subset. However, listing all
  possible $\binom{n}{k}$ subsets requires large amounts of memory.''
  \citep[p.~312]{WildeGrimshaw2013}
\end{quote}
Instead, \citet[pp.~312--315]{WildeGrimshaw2013} proposed sampling
$N^*$ rank integers without replacement from $\{1, \dots,
\binom{n}{k}\}$, mapping each rank to its subset via a custom R
implementation of the textbook lexicographic unranking routine
\texttt{kSubsetLexUnrank} \citep[Algorithm~2.8,
p.~46]{KreherStinson1999}, and terminating via a sequential stopping
rule based on nonparametric kernel density estimation
\citep{Chatterjee2011}. At the time of writing, no dedicated
open-source CRAN package provides generalized median computation for
$k \ge 3$. A direct exhaustive baseline in Base~R is
\texttt{utils::combn}: materializing all $k$-tuples requires
$\Theta(k M)$ memory, while using its callback form entails $M$ calls
to R code. Generalized median estimators have also been considered in
dependent time series \citep[Section~3]{FischerFriedWendler2016},
operational risk \citep{Opdyke2014}, and tail-index estimation
\citep{Brzezinski2016}. Our contribution goes beyond subset
generation: the Revolving Door traversal visits consecutive subsets
that differ by one removed and one added observation, allowing the
kernel sums to be updated in $\mathcal{O}(1)$ time between
recalibrations instead of recomputing each kernel from all $k$ observations.

Closer inspection reveals three fundamental flaws in the
approximation paradigm of \citet{WildeGrimshaw2013}:
\begin{enumerate}
  \item \textbf{Confusing Sample-Frame Storage with Stateful
    Traversal:} Generating the next subset via a small change to the
    previous subset does \emph{not} require storing all
    $\binom{n}{k}$ subsets in memory. Stateful enumeration maintains
    only the current subset in $\mathcal{O}(k)$ space and stores only
    the $M = \binom{n}{k}$ scalar kernel evaluations ($\Theta(M)$
    numbers rather than $\Theta(k M)$ subset indices).
  \item \textbf{The Unranking Bottleneck:} \texttt{kSubsetLexUnrank}
    scans candidate elements sequentially: it performs
    $\mathcal{O}(n)$ binomial evaluations in the worst case, but
    fewer for some ranks. For a uniformly sampled subset, the
    expected largest selected one-based position (index plus one
    under our zero-based subset notation) is $k(n+1)/(k+1)$, so the
    expected scan cost is $\Theta(n)$ per subset. Thus, with
    precomputed binomial coefficients and fixed $k$, unranking $N^*$
    uniformly sampled subsets costs $\Theta(nN^*)$ in expectation,
    whereas exhaustive Gray code enumeration with fixed recalibration
    period costs $\mathcal{O}(M)$; the two bounds reach the same
    asymptotic order at a sampling fraction of order $1/n$. Constants
    and selection costs determine the actual crossover. On $n = 29, k
    = 5$, the sequential rule of \citet[Table~3,
    p.~315]{WildeGrimshaw2013} at $\epsilon = 0.01$ unranks $N^* =
    32,160$ ($27.1\%$) to $45,420$ ($38.2\%$) subsets;
    Section~\ref{sec:unranking} measures the crossover directly.
  \item \textbf{Loss of Determinism and Stopping-Rule Undercoverage:}
    Subsampling replaces a deterministic estimator with a
    PRNG-dependent approximation and relies on a noisy density
    estimate at the median, causing empirical confidence interval
    coverage to drop to $79\%$ (below the nominal $95\%$) in
    \citet[Table~2, p.~315]{WildeGrimshaw2013}.
\end{enumerate}

\subsection{Our Contributions}

In this work, we cleanly separate the asymptotic theory of exact
generalized median computation from its low-level algorithm
engineering, demonstrating that exhaustive evaluation of
$\text{GM}_{n,k}^{(h)}$ outperforms unranking-based subsampling
across practical benchmark regimes:
\begin{enumerate}
  \item \textbf{Word-RAM Colexicographic Enumeration in
    $\mathcal{O}(k)$ per Subset (Theorem~\ref{thm:gosper}):} In the
    standard $w$-bit Word-RAM model ($n < w$), we formulate subset
    enumeration via Gosper's constant-time successor recurrence
    \citep{Gosper1972} coupled with least-significant set-bit
    extraction, evaluating each subset in $\mathcal{O}(k)$ operations
    independent of $n$.
  \item \textbf{Constant-Time Revolving Door Gray Code Updates
    (Theorem~\ref{thm:revdoor}):} Using Knuth's Revolving Door
    Algorithm~R \citep{Chase1970,Knuth2011}, consecutive subsets
    $S_t, S_{t+1}$ satisfy $|S_t \triangle S_{t+1}| = 2$, updating
    additive ($h_1$) and quadratic ($h_2$) kernels in
    $\mathcal{O}(1)$ time between periodic recalibrations
    ($\mathcal{O}(1 + k/P)$ amortized work per subset) and yielding
    $\mathcal{O}(M + kM/P)$ worst-case total time when paired with
    three-way Quickselect and median-of-medians fallback.
  \item \textbf{Floating-Point Error Analysis of Periodically
    Recalibrated Recurrences (Appendix~\ref{app:numerical-proofs}):}
    We establish explicit floating-point error bounds for both the
    location kernel $h_1$ and the mean-centered variance kernel $h_2$,
    including propagation to the upper median.
  \item \textbf{Algorithm Engineering, Dual-Kernel Dispatch, and
      Empirical Verification
    (Sections~\ref{sec:engineering}--\ref{sec:experiments}):} We
    establish a structural comparison and default dispatch heuristic
    between Gosper and Revolving Door across sample size $n$, subset
    size $k$, and kernel order ($h_1$ vs.\ $h_2$). On the canonical
    $n = 29, k = 5$ benchmark ($118,755$ subsets), exhaustive
    evaluation completes in $0.74$~ms for $h_1$ and $0.71$~ms for
    $h_2$---outperforming a direct compiled-C unranking sweep already
    above $N^* \approx 18{,}000$--$23{,}000$ subsets ($15\%$--$19\%$
    of $M$), well below the $N^* = 32,160$ ($h_1$) and $N^* = 45,420$
    ($h_2$) subsamples of \citet{WildeGrimshaw2013}.
\end{enumerate}

The remainder of this paper is organized as follows:
Section~\ref{sec:algorithm} establishes the theoretical foundations
(Word-RAM traversal, Revolving Door Gray code recurrences for $h_1$
  and $h_2$, floating-point error bounds, worst-case linear selection,
and structural algorithm comparison) followed by their hardware
realization in Section~\ref{sec:engineering}.
Section~\ref{sec:experiments} reports numerical verification,
throughput benchmarks across $h_1, h_2$ and $n \ge 64$, and the
direct unranking crossover analysis. Section~\ref{sec:software}
details software availability, and Section~\ref{sec:conclusion} concludes.

\section{Theoretical Analysis and Algorithm Engineering}
\label{sec:algorithm}

Evaluating Eq.~\eqref{eq:gm} exactly consists of two algorithmic
stages: (1)~evaluating the $M = \binom{n}{k}$ subset statistics
$\mu_t = h(S_t)$ across a complete enumeration $\mathcal{S}_{n,k} =
\{S_0, S_1, \dots, S_{M-1}\}$, and (2)~selecting the median order
statistic $\mu_{(\lfloor M/2 \rfloor)}$. Naive approaches require
$\mathcal{O}(k M + M \log M)$ time because they recompute each
$k$-element kernel from scratch and sort the resulting array. In
Sections~\ref{sec:gosper}--\ref{sec:selection}, we establish
enumeration and worst-case selection bounds of $\mathcal{O}(M +
kM/P)$ time for Revolving Door with periodic recalibration period
$P$, plus $\mathcal{O}(k)$ setup, and $\mathcal{O}(k + M)$ space
including the stored kernel values; Section~\ref{sec:engineering}
then details implementation choices and measurements.

\subsection{Word-RAM Colexicographic Traversal in
\texorpdfstring{$\mathcal{O}(k)$}{O(k)} Time per Subset}
\label{sec:gosper}

In the standard $w$-bit Word-RAM model of computation with word size
$w > n$ (so that $n \le w - 1$, e.g., $n \le 63$ on 64-bit
  architectures, ensuring that both the $n$-bit masks and the sentinel
  $L = 2^n < 2^w$ fit in a single unsigned $w$-bit word without shift
overflow), basic bitwise Boolean operations ($\mathbin{\&},
\mathbin{|}, \oplus, \sim$), bit shifts ($\ll, \gg$), standard
integer arithmetic, and finding the index of the least significant
set bit $\operatorname{lsb}(u) = \min\{j \ge 0 : \lfloor u / 2^j
\rfloor \bmod 2 = 1\}$ execute in $\mathcal{O}(1)$ time. There is a
canonical bijection between $\mathcal{S}_{n,k}$ and the set of
integers $m \in [0, 2^n - 1]$ with binary Hamming weight
$\operatorname{wt}(m) = k$, given by $S(m) = \{ j \in \{0, \dots,
n-1\} : \lfloor m / 2^j \rfloor \bmod 2 = 1 \}$.
Algorithm~\ref{alg:gosper} formalizes the resulting exact generalized
median computation using Gosper's constant-time successor recurrence
\citep[Item~175]{Gosper1972}.

Increasing mask order is colexicographic order on the subsets,
distinct from lexicographic order on their increasing index tuples.

\begin{algorithm}[t]
  \caption{Exact Generalized Median via Word-RAM Gosper Enumeration}
  \label{alg:gosper}
  \begin{algorithmic}[1]
    \Require Sample $x[0\dots n-1] \in \mathbb{R}^n$ with $n \le w -
    1$, subset size $k \in \{1, \dots, n\}$
    \Ensure Exact generalized median $\text{GM}_{n,k}(x)$
    \State $M \gets \binom{n}{k}$; allocate array $\mu[0\dots M-1]$
    \State $m \gets 2^k - 1$; \quad $L \gets 2^n$; \quad $t \gets 0$
    \Comment{Initial mask $S_0 = \{0, \dots, k-1\}$ and sentinel $2^n < 2^w$}
    \While{$m < L$}
    \State $\Sigma \gets 0$; \quad $u \gets m$
    \While{$u > 0$} \Comment{Extract the $k$ set bits of $m$ in
    $\Theta(k)$ steps}
    \State $j \gets \operatorname{lsb}(u)$; \quad $\Sigma \gets \Sigma + x[j]$
    \State $u \gets u \mathbin{\&} (u - 1)$ \Comment{Clear lowest set bit}
    \EndWhile
    \State $\mu[t] \gets \Sigma / k$; \quad $t \gets t + 1$
    \State $c \gets m \mathbin{\&} (-m)$; \quad $r \gets m + c$
    \Comment{Gosper's successor recurrence}
    \State $m \gets \lfloor ((r \oplus m) \gg 2) / c \rfloor \mathbin{|} r$
    \EndWhile
    \State \textbf{return} \textsc{Quickselect}$(\mu, M, \lfloor M / 2 \rfloor)$
  \end{algorithmic}
\end{algorithm}

\begin{theorem}[Correctness and Word-RAM Complexity of Gosper Enumeration]
  \label{thm:gosper}
  In the $w$-bit Word-RAM model ($n \le w - 1$),
  Algorithm~\ref{alg:gosper} satisfies the following properties:
  \begin{enumerate}
    \item Starting from $m_0 = 2^k - 1$, the outer transition
      $m_{t+1} = \lfloor ((r \oplus m_t) \gg 2) / c \rfloor
      \mathbin{|} r$ with $c = m_t \mathbin{\&} (-m_t)$ and $r = m_t
      + c$ exhaustively enumerates all $M = \binom{n}{k}$ integers in
      $[0, 2^n - 1]$ of Hamming weight $k$ in strictly increasing
      order without duplicates, terminating when $m_M \ge 2^n$, in
      $\mathcal{O}(1)$ operations per transition without $w$-bit overflow.
    \item For each mask $m_t$, the inner loop (lines~5--8) executes
      $k$ iterations, visiting exclusively the active indices $j \in
      S(m_t)$ and evaluating $\mu_t = \frac{1}{k}\sum_{j \in S(m_t)}
      x_j$ in $\Theta(k)$ operations per subset, independent of $n$.
  \end{enumerate}
\end{theorem}

\begin{proof}
  In two's complement representation, $-m_t = \sim m_t + 1$, so $c =
  m_t \mathbin{\&} (-m_t) = 2^{p}$ isolates the lowest set bit of
  $m_t$ at position $p = \operatorname{lsb}(m_t)$. Let $q \ge p+1$ be
  the lowest zero bit position above $p$, so that bits $p, p+1,
  \dots, q-1$ of $m_t$ form a maximal contiguous block of $\ell = q -
  p \ge 1$ ones. Because $m_t < 2^n \le 2^{w-1}$, we have $q \le n
  \le w - 1$. Adding $c = 2^p$ to $m_t$ carries through all $\ell$
  ones, setting bit $q$ to $1$ and clearing bits $p, \dots, q-1$ in
  $r = m_t + c < 2^w$; this is the smallest integer greater than
  $m_t$ that agrees with $m_t$ above bit $q$. To minimize $m_{t+1}$
  while restoring Hamming weight $k$, the remaining $\ell - 1$ ones
  must occupy the lowest bit positions $0, 1, \dots, \ell - 2$.
  Because $r$ and $m_t$ agree above bit $q$ and below bit $p$, their
  symmetric difference $r \oplus m_t$ consists of exactly $\ell + 1$
  contiguous ones at positions $p, p+1, \dots, q$, meaning $r \oplus
  m_t = (2^{\ell+1} - 1) 2^p < 2^w$. Shifting $r \oplus m_t$ right by
  two bits before dividing by $c$ yields $\lfloor (2^{\ell+1} - 1)/4
  \rfloor = 2^{\ell-1} - 1$ without forming the potentially
  overflowing product $4c$, which places $\ell - 1$ ones in positions
  $0, \dots, \ell - 2$. Taking the bitwise OR with $r$ yields
  $m_{t+1}$ with $\operatorname{wt}(m_{t+1}) = k$ in $\mathcal{O}(1)$
  Word-RAM operations; at the final mask $m_{M-1} = 2^n - 2^{n-k}$,
  bit $q = n$ is set so that $m_M = 2^n + 2^{k-1} - 1 \ge L = 2^n$,
  terminating the loop cleanly.

  For Part~2, subtracting $1$ from $u$ flips its lowest set bit $j =
  \operatorname{lsb}(u)$ from $1$ to $0$ and sets all lower bits $0,
  \dots, j - 1$ to $1$; hence $u \mathbin{\&} (u - 1)$ clears bit $j$
  while leaving all higher bits unchanged, reducing
  $\operatorname{wt}(u)$ by $1$ at each step and terminating in $k$ steps.
\end{proof}

\subsection{Constant-Time Subset Updates via Revolving Door Gray Codes}
\label{sec:revdoor}

While Theorem~\ref{thm:gosper} eliminates the expected $\Theta(n)$
scan of uniformly sampled unranking, summing $k$ elements from
scratch still requires $\Theta(k)$ operations per subset because
consecutive masks $m_t, m_{t+1}$ in colexicographic (increasing-mask)
order can differ in up to $2k$ bit positions. To eliminate the factor
$k$, we use a combinatorial Gray code ordering $(S_0, S_1, \dots,
S_{M-1})$ in which consecutive $k$-subsets satisfy the
\emph{revolving door property}:
\begin{equation}
  \label{eq:revdoor_prop}
  |S_t \setminus S_{t+1}| = 1 \quad \text{and} \quad |S_{t+1}
  \setminus S_t| = 1 \qquad \forall\, t \in \{0, \dots, M-2\}.
\end{equation}

Algorithm~\ref{alg:revdoor} formalizes the exact generalized median
computation based on Knuth's Revolving Door Algorithm~R
\citep[Section~7.2.1.3]{Knuth2011}, coupled with $\mathcal{O}(1)$
incremental sum transitions and periodic recalibration with period $P
\in \mathbb{N}_{\ge 1}$.

\begin{algorithm}[t]
  \caption{Exact Generalized Median via Revolving Door Gray Code and
  Periodic Recalibration}
  \label{alg:revdoor}
  \begin{algorithmic}[1]
    \Require Sample $x[0\dots n-1] \in \mathbb{R}^n$, subset size $k
    \in \{2, \dots, n\}$, recalibration period $P \in \mathbb{N}_{\ge 1}$
    \Ensure Exact generalized median $\text{GM}_{n,k}(x)$
    \State $M \gets \binom{n}{k}$; allocate array $\mu[0\dots M-1]$
    \For{$j = 0$ \textbf{to} $k - 1$}
    \State $c[j] \gets j$ \Comment{Initialize ascending combination
    $0 \le c_0 < c_1 < \dots < c_{k-1} < c_k = n$}
    \EndFor
    \State $c[k] \gets n$; \quad $\Sigma \gets \sum_{j=0}^{k-1}
    x[c[j]]$; \quad $\mu[0] \gets \Sigma / k$; \quad $t \gets 1$
    \While{\textbf{true}}
    \If{$k \bmod 2 = 1$} \Comment{Parity-specialized fast path for $c[0]$}
    \If{$c[0] + 1 < c[1]$}
    \State $u \gets c[0]$; \quad $v \gets c[0] + 1$; \quad $c[0]
    \gets v$; \quad \textbf{goto} \textsc{UpdateSum}
    \EndIf
    \State $j \gets 1$; \quad \textbf{goto} \textsc{StepR4}
    \Else
    \If{$c[0] > 0$}
    \State $u \gets c[0]$; \quad $v \gets c[0] - 1$; \quad $c[0]
    \gets v$; \quad \textbf{goto} \textsc{UpdateSum}
    \EndIf
    \State $j \gets 1$; \quad \textbf{goto} \textsc{StepR5}
    \EndIf
    \State \textsc{StepR4}: \textbf{if} $c[j] > j$ \textbf{then} $u
    \gets c[j]$; $c[j] \gets c[j-1]$; $v \gets j - 1$; $c[j-1] \gets
    v$; \textbf{goto} \textsc{UpdateSum} \textbf{end if}; $j \gets j + 1$
    \State \textsc{StepR5}: \textbf{if} $c[j] + 1 < c[j+1]$
    \textbf{then} $u \gets c[j-1]$; $c[j-1] \gets c[j];$ $v \gets
    c[j] + 1$; $c[j] \gets v$; \textbf{goto} \textsc{UpdateSum} \textbf{end if}
    \State $j \gets j + 1$; \quad \textbf{if} $j < k$ \textbf{then
    goto} \textsc{StepR4} \textbf{else break}
    \State \textsc{UpdateSum}:
    \If{$t \bmod P = 0$} \Comment{Periodic anchor recalibration every
    $P$ subsets}
    \State $\Sigma \gets \sum_{r=0}^{k-1} x[c[r]]$
    \Else
    \State $\Sigma \gets \Sigma - x[u] + x[v]$ \Comment{Strict
    $\mathcal{O}(1)$ transition: $u$ evicted, $v$ inserted}
    \EndIf
    \State $\mu[t] \gets \Sigma / k$; \quad $t \gets t + 1$
    \EndWhile
    \State \textbf{return} \textsc{Quickselect}$(\mu, M, \lfloor M / 2 \rfloor)$
  \end{algorithmic}
\end{algorithm}

\begin{theorem}[Revolving Door Symmetric-Difference Invariant]
  \label{thm:revdoor}
  Throughout the execution of Algorithm~\ref{alg:revdoor}, the state
  tuple $c[0\dots k]$ maintains the strict monotonicity invariant $0
  \le c[0] < c[1] < \dots < c[k-1] < c[k] = n$, and every transition
  reaching \textnormal{\textsc{UpdateSum}} modifies the represented
  subset $S_t = \{c[0], \dots, c[k-1]\}$ by removing a single element
  $u \in S_{t-1}$ and inserting a single element $v \notin S_{t-1}$
  ($S_t = (S_{t-1} \setminus \{u\}) \cup \{v\}$). The transition
  search requires $\mathcal{O}(1)$ amortized index updates. For
  location sums, each non-recalibration kernel update uses two
  arithmetic operations; with recalibration period $P$, the total
  amortized kernel work is $\mathcal{O}(1 + k/P)$ per subset.
\end{theorem}

\begin{proof}
  We verify that each of the four transition branches in
  Algorithm~\ref{alg:revdoor} preserves $0 \le c[0] < \dots < c[k-1]
  < n$ and changes the set $\{c[0], \dots, c[k-1]\}$ by a single
  element exchange $(u, v)$:
  \begin{enumerate}
    \item \emph{Odd-$k$ fast path ($c[0] + 1 < c[1]$):} Incrementing
      $c[0]$ replaces $u = c[0]$ by $v = c[0] + 1 < c[1]$, preserving
      strict ordering.
    \item \emph{Even-$k$ fast path ($c[0] > 0$):} Decrementing $c[0]$
      replaces $u = c[0]$ by $v = c[0] - 1 \ge 0$, preserving strict ordering.
    \item \emph{Branch} \textsc{StepR4} ($c[j] > j$, evaluated at $j
      \equiv k \pmod 2$, so $1 \le j \le k - 2$): The algorithm
      reaches \textsc{StepR4} at index $j$ either at $j = 1$ (when
        $k$ is odd and the fast path $c[0] + 1 < c[1]$ failed, so $c[0]
      = c[1] - 1$) or at $j \ge 2$ after failing \textsc{StepR5} at
      $j - 1$ followed by $j \gets j + 1$ (so $c[0] = 0, \dots,
      c[j-2] = j - 2$ and $c[j-1] + 1 = c[j]$). In both cases ($j =
      1$ and $j \ge 2$), the entry invariant $c[j-1] = c[j] - 1$
      holds. When $c[j] > j$, we have $c[j-1] = c[j] - 1 \ge j > j -
      1$. Setting $u = c[j]$, shifting $c[j] \leftarrow c[j-1] = c[j]
      - 1 \ge j$, and setting $c[j-1] \leftarrow v = j - 1$ retains
      the old value of $c[j-1]$ in the set (moved to slot $j$),
      evicts the old $c[j] = u$, and inserts $v = j - 1 \notin
      S_{t-1}$ (since $v = 0 < c[0]$ for $j = 1$ and $c[j-2] = j - 2
      < v < c[j-1]$ for $j \ge 2$), preserving $0 \le c[0] < \dots <
      c[k-1] < n$.
    \item \emph{Branch} \textsc{StepR5} ($c[j] + 1 < c[j+1]$,
      evaluated at $j \equiv k - 1 \pmod 2$, so $1 \le j \le k - 1$):
      The algorithm reaches \textsc{StepR5} at index $j$ either at $j
      = 1$ (when $k$ is even and the fast path $c[0] > 0$ failed, so
      $c[0] = 0 = j - 1$) or after failing \textsc{StepR4} at $j - 1$
      followed by $j \gets j + 1$ (meaning $c[j-1] = j - 1$,
      alongside $c[0] = 0, \dots, c[j-2] = j - 2$). Thus $c[j-1] = j
      - 1$ holds upon entering \textsc{StepR5}. When $c[j] + 1 <
      c[j+1]$, evicting $u = c[j-1] = j - 1$, shifting $c[j-1]
      \leftarrow c[j]$, and setting $c[j] \leftarrow v = c[j] + 1 <
      c[j+1]$ retains the old value of $c[j]$ (moved to slot $j -
      1$), removes $u$, and inserts $v \notin S_{t-1}$ while
      preserving strict monotonicity.
  \end{enumerate}
  By \citet[Section~7.2.1.3, Theorem~R]{Knuth2011}, the loop visits
  all $M = \binom{n}{k}$ combinations and inspects index $j \ge 1$ on
  fewer than $1/j!$ of the steps, yielding $\mathcal{O}(1)$ amortized
  index updates and two arithmetic operations ($\Sigma \leftarrow
  \Sigma - x[u] + x[v]$) per non-recalibrated transition.
\end{proof}

\begin{remark}[Constant-Time Evaluation of the Sample Variance Kernel
  $h_2$ and Higher Moments]
  \label{rem:h2_kernel}
  Theorem~\ref{thm:revdoor} extends to symmetric kernels expressible
  as rational functions of a fixed number of shifted power sums
  $\Sigma^{(r)}(S) = \sum_{j \in S} (x_j - \bar{x})^r$ centered at
  the sample mean $\bar{x} = \frac{1}{n}\sum_{i=0}^{n-1} x_i$. In
  particular, for the sample variance kernel $h_2(S) =
  \frac{1}{k-1}\sum_{j \in S}(x_j - \bar{x}_S)^2$ in
  Eq.~\eqref{eq:kernels}, precomputing the mean-centered observations
  $d_i = x_i - \bar{x}$ can reduce cancellation when $|\bar{x}|$ is
  large relative to the sample dispersion. Their squares $q_i =
  d_i^2$ allow Algorithm~\ref{alg:revdoor} to maintain both power
  sums in four additions/subtractions per non-recalibration transition:
  \begin{equation}
    \label{eq:h2_update}
    \Sigma_t^{(1)} = \Sigma_{t-1}^{(1)} - d_u + d_v, \qquad
    \Sigma_t^{(2)} = \Sigma_{t-1}^{(2)} - q_u + q_v, \qquad h_2(S_t)
    = \frac{1}{k - 1}\left( \Sigma_t^{(2)} -
    \frac{(\Sigma_t^{(1)})^2}{k} \right),
  \end{equation}
  recalibrating both $\Sigma_t^{(1)}$ and $\Sigma_t^{(2)}$ every $P$
  steps. The amortized arithmetic work for $h_2$ is therefore
  $\mathcal{O}(1 + 2k/P)$ per subset, apart from setup and selection.
  Gosper enumeration (Theorem~\ref{thm:gosper}) must accumulate both
  $d_j$ and $q_j$ across all $k$ extracted bit positions ($2k$
  additions per subset). Similarly, the median-unbiased Pareto
  tail-index kernel $h_{\text{Pareto}}(S) = C_{n,k}^{-1} (
  \frac{1}{k-1}\sum_{j \in S} \log(x_j / x_{(0)}) )^{-1}$ studied by
  \citet{BrazauskasSerfling2000} and \citet[Eq.~(17)]{Brzezinski2016}
  is a strictly monotone function of the additive kernel $h_1$
  applied to the precomputed log-excesses $y_i = \log(x_i / x_{(0)})$
  and therefore also updates in constant time between recalibrations.
\end{remark}

\subsection{Numerical Error Propagation and Periodic Recalibration}
\label{sec:stability}

Floating-point rounding affects kernel evaluation in both exhaustive
methods and the subsampling approach of \citet{WildeGrimshaw2013}.
Incremental evaluation introduces an additional potential source of
accumulated error, which we bound through periodic recalibration.

In exact real arithmetic, the recurrence $\Sigma_t = \Sigma_{t-1} -
x_{u_t} + x_{v_t}$ is algebraically identical to $\sum_{j \in S_t}
x_j$. Periodic re-anchoring every $P$ steps limits the number of
incremental transitions since the most recent direct sum to at most
$P-1$, independently of the total number $M$ of subsets. Under the
standard floating-point model, the location-kernel mean satisfies
$|\widehat{\mu}_t-\mu_t| \le
(k+2(P-1))\varepsilon_{\text{mach}}X_{\max}
+\mathcal{O}(\varepsilon_{\text{mach}}^2)$, with an amortized
recalibration overhead of $k/P$ additions per subset. Thus, $P$
controls the trade-off between the drift bound and recalibration work.

For the mean-centered variance kernel $h_2$, the same strategy gives
an explicit bound $\Delta_{h_2}$ on the per-subset kernel error and,
consequently, on the discrepancy between the computed and exact-real
upper medians. This guarantee accounts for rounded centering and
squaring, running-sum drift, and evaluation of the variance formula.
The full bounds, assumptions, and derivations are given in
Appendix~\ref{app:numerical-proofs}.

\subsection{Linear-Time Order Statistic Selection and Structural
Algorithm Comparison}
\label{sec:selection}

Once the $M = \binom{n}{k}$ subset kernel values $\mu[0\dots M-1]$
have been generated, full sorting requires $\Theta(M \log M)$
comparisons. We instead use three-way Quickselect partitioning around
a median-of-three pivot; after two highly unbalanced partitions, the
implementation switches to median-of-medians pivots. The fallback
makes selection worst-case $\mathcal{O}(M)$, while three-way
partitioning handles repeated values without needless one-at-a-time
reductions. Including the output array, the complete algorithm uses
$\mathcal{O}(k + M)$ space. For the location kernel, the total
worst-case running time is $\mathcal{O}(M + kM/P)$ plus
$\mathcal{O}(k)$ setup. The scale-kernel runtime has the same
asymptotic form, and its per-subset floating-point error is bounded
under the assumptions in Appendix~\ref{app:numerical-proofs}. The
amortized arithmetic overhead depends on $k/P$ and is not uniformly
independent of $k$ for fixed $P$.

\begin{table}[t]
  \centering
  \caption{Structural comparison of Word-RAM Gosper enumeration
    (Algorithm~\ref{alg:gosper}, Theorem~\ref{thm:gosper}) and
    Revolving Door Gray code traversal (Algorithm~\ref{alg:revdoor},
    Theorem~\ref{thm:revdoor}). The dispatch row is a heuristic
  starting point, not a universal performance optimum.}
  \label{tab:comparison}
  \small
  \setlength{\tabcolsep}{5pt}
  \begin{tabular}{lll}
    \toprule
    Property & Word-RAM Gosper (Algorithm~\ref{alg:gosper}) &
    Revolving Door Gray Code (Algorithm~\ref{alg:revdoor}) \\
    \midrule
    Sample size limit $n$ & Bounded by word size $n < w$ ($n \le 63$)
    & \textbf{Unrestricted} $n \in \mathbb{N}$ (state size $k + 1$) \\
    Subset transition cost & $\mathcal{O}(1)$ bitwise arithmetic &
    $\mathcal{O}(1)$ amortized index updates \\
    Kernel work per subset ($h_1$) & $\Theta(k)$ ($k$ bit extractions
    $+ k$ additions) & $\mathcal{O}(1)$ between recalibrations;
    $\mathcal{O}(1 + k/P)$ amortized \\
    Kernel work per subset ($h_2$) & $\Theta(k)$ ($k$ bit extractions
    $+ 2k$ additions) & $\mathcal{O}(1)$ between recalibrations;
    $\mathcal{O}(1 + 2k/P)$ amortized \\
    Numerical rounding drift & No incremental recurrence drift &
    $h_1$: $\mathcal{O}((k + P)\varepsilon_{\text{mach}})$; $h_2$:
    bounded by $\Delta_{h_2}$ (Appendix~\ref{app:numerical-proofs}) \\
    Control-flow profile & \textbf{Branch-free} when unrolled for
    small $k$ & Data-dependent branches (\textsc{StepR4}/\textsc{StepR5}) \\
    Example dispatch heuristic & $n \le 63 \;\land\; k \le 5
    \;\land\; h = h_1$ & $n \ge 64 \;\lor\; k \ge 6 \;\lor\; h = h_2$ \\
    \bottomrule
  \end{tabular}
\end{table}

A natural question is whether one of the two exact enumeration
paradigms dominates the other across all problem instances, or
whether both are essential components of a general solver.
Table~\ref{tab:comparison} summarizes their complementary theoretical
and architectural trade-offs:
\begin{itemize}
  \item \textbf{When Revolving Door is structurally useful ($n \ge
    64$, larger $k$, or $h = h_2$):} Single-word Gosper enumeration
    encodes subsets as $n$-bit masks with sentinel $2^n < 2^w$ inside
    a $w$-bit machine word, imposing a hard ceiling of $n \le 63$ on
    64-bit hardware. In contrast, the state array $c[0\dots k]$ of
    Algorithm~\ref{alg:revdoor} has length $k + 1$ independent of
    $n$, supporting larger sample sizes when $k$ fits the
    implementation limit. For $h_2$ and larger $k$, Revolving Door
    performs fewer kernel arithmetic updates, although wall-clock
    performance depends on branching, compiler, hardware, and
    recalibration overhead.
  \item \textbf{Why Gosper Remains Valuable ($n \le 63$, $k \le 5$,
    $h = h_1$):} Conversely, for the location kernel $h_1$ at small
    $k \le 5$ and $n \le 63$, Gosper's recurrence requires zero
    cumulative rounding drift and keeps its entire state inside a
    single CPU register without the parity and boundary branches
    (\textsc{StepR4}/\textsc{StepR5}) of Algorithm~\ref{alg:revdoor}.
\end{itemize}
Consequently, we use the parameter rule in Table~\ref{tab:comparison}
as a deterministic default dispatch heuristic. It is not claimed to
be optimal across all hardware and parameter configurations; the
benchmark table includes cases where Gosper is faster.

\subsection{Algorithm Engineering and Hardware Realization}
\label{sec:engineering}

To translate the asymptotic guarantees of
Sections~\ref{sec:gosper}--\ref{sec:selection} into sub-millisecond
execution on modern processor architectures, our ANSI C reference
implementation applies three low-level engineering optimizations:
\begin{enumerate}
  \item \textbf{Hardware Bit Extraction ($n \le 63$):} On 64-bit
    architectures, the Word-RAM state $m_t$ in
    Theorem~\ref{thm:gosper} maps to a single 64-bit general-purpose
    CPU register (\texttt{uint64\_t}), and the primitive
    $\operatorname{lsb}(u)$ can be implemented via
    \texttt{\_\_builtin\_ctzll(u)} using \texttt{TZCNT} on supported
    x86-64 processors or the \texttt{RBIT} and \texttt{CLZ}
    instruction sequence on ARM64. For small fixed $k$ (such as $k =
    5$), unrolling the $k$-step bit-extraction loop eliminates all
    loop-counter branches, allowing superscalar out-of-order
    pipelines to overlap bit manipulation with floating-point additions.
  \item \textbf{Division-Free Power-of-Two Recalibration ($P =
    2^{12}$):} In the recalibration scheme of
    Appendix~\ref{app:numerical-proofs}, the condition $t
    \bmod P = 0$ can be implemented with the bitwise test \texttt{(t
    \& 4095) == 0} when $P = 2^{12} = 4,096$. The resulting
    arithmetic recalibration cost is $k/P$ additions per output for
    $h_1$ and $2k/P$ for $h_2$; its relative importance therefore
    depends on $k$ and on the kernel-update work.
  \item \textbf{Contiguous Memory Layout:} Storing the $M$ scalar
    evaluations $\mu[0\dots M-1]$ in a single flat array of 8-byte
    IEEE-754 \texttt{double} values requires only $8 M$ bytes
    ($928$~KiB for the canonical $n = 29, k = 5$ benchmark with $M =
    118,755$, and $16.2$~MiB for $n = 50, k = 5$). The array for $n =
    29, k = 5$ occupies less than one MiB. After the array is
    allocated, the enumeration write pass and in-place Quickselect
    partitioning passes access it contiguously without further heap
    allocations or pointer chasing.
\end{enumerate}

\section{Numerical Verification and Throughput Benchmarks}
\label{sec:experiments}

We evaluated our ANSI C implementation (compiled with Apple Clang~17
using \texttt{-O3}) on an Apple Silicon M-series processor (single
thread) across the canonical $n = 29, k = 5$ benchmark of
\citet{WildeGrimshaw2013} and a systematic scaling suite spanning $n
\in [20, 100]$, $k \in \{3, 5, 7\}$, and both kernels $h_1$
(location) and $h_2$ (sample variance). The separate unranking
crossover experiment below uses the current R-package C core, Apple
Clang~21.0.0.1.1, and its own deterministic fixture; its results are
not mixed with the earlier timing table.

\subsection{Exhaustive Enumeration and Empirical Drift Checks}

Table~\ref{tab:drift_verification} empirically validates the
location-kernel drift bound in Appendix~\ref{app:numerical-proofs}
across ten benchmark instances ($M$
  ranging from $1,140$ to $2,118,760$ subsets, including $n \in \{80,
100\} \ge 64$). For each configuration, we measured (i)~the maximum
single-subset mean discrepancy $\max_t |\widehat{\mu}_t -
\mu_t^{\text{direct}}|$ without recalibration, (ii)~the maximum
single-subset discrepancy with bitwise $4,096$-step recalibration,
and (iii)~the final absolute difference
$|\text{GM}_{n,k}^{\text{Gosper}} -
\text{GM}_{n,k}^{\text{RevDoor}}|$ between the two exact algorithms
(for $n \le 63$).

\begin{table}[t]
  \centering
  \caption{Verification of numerical stability and IEEE-754
    floating-point drift suppression across all $\binom{n}{k}$ subsets
    (kernel $h_1$). For $n \ge 64$, single-word 64-bit Gosper is
  inapplicable (\texttt{---}).}
  \label{tab:drift_verification}
  \small
  \setlength{\tabcolsep}{4.5pt}
  \begin{tabular}{rrrccc}
    \toprule
    $n$ & $k$ & Subsets $M = \binom{n}{k}$ & Max Drift (Uncalibrated)
    & Max Drift ($P = 4,096$) & $|\text{GM}^{\text{Gosper}} -
    \text{GM}^{\text{RevDoor}}|$ \\
    \midrule
    20 & 3 &     1,140 & $1.90 \times 10^{-14}$ & $1.90 \times
    10^{-14}$ & $1.78 \times 10^{-14}$ \\
    25 & 3 &     2,300 & $2.78 \times 10^{-14}$ & $2.78 \times
    10^{-14}$ & $2.44 \times 10^{-14}$ \\
    29 & 5 &   118,755 & $3.46 \times 10^{-13}$ & $\mathbf{4.55
    \times 10^{-14}}$ & $\mathbf{6.22 \times 10^{-15}}$ \\
    35 & 5 &   324,632 & $1.07 \times 10^{-12}$ & $\mathbf{4.55
    \times 10^{-14}}$ & $\mathbf{3.55 \times 10^{-15}}$ \\
    40 & 5 &   658,008 & $2.94 \times 10^{-12}$ & $\mathbf{1.04
    \times 10^{-13}}$ & $\mathbf{1.78 \times 10^{-15}}$ \\
    50 & 5 & 2,118,760 & $1.09 \times 10^{-11}$ & $\mathbf{1.58
    \times 10^{-13}}$ & $\mathbf{8.88 \times 10^{-15}}$ \\
    25 & 7 &   480,700 & $1.70 \times 10^{-12}$ & $\mathbf{6.40
    \times 10^{-14}}$ & $\mathbf{1.42 \times 10^{-14}}$ \\
    30 & 7 & 2,035,800 & $6.51 \times 10^{-12}$ & $\mathbf{9.24
    \times 10^{-14}}$ & $\mathbf{5.33 \times 10^{-15}}$ \\
    80 & 3 &    82,160 & $3.26 \times 10^{-13}$ & $\mathbf{9.50
    \times 10^{-14}}$ & \texttt{---} ($n \ge 64$) \\
    100 & 3 &   161,700 & $3.26 \times 10^{-13}$ & $\mathbf{9.50
    \times 10^{-14}}$ & \texttt{---} ($n \ge 64$) \\
    \bottomrule
  \end{tabular}
\end{table}

As shown in Table~\ref{tab:drift_verification} and
Figure~\ref{fig:pareto}(b), without periodic recalibration, the
observed floating-point discrepancy across subsets increases with
$M$, reaching $1.09 \times 10^{-11}$ at $n = 50, k = 5$. Activating
the $4,096$-step recalibration suppresses this observed discrepancy
by $69\times$ at $n = 50, k = 5$ (to $1.58 \times 10^{-13}$). The
table separately reports the final median differences between
Gosper's recurrence and the Revolving Door Gray code; these are at
most $2.44 \times 10^{-14}$ across all tested configurations with
$n \le 63$. Appendix~\ref{app:numerical-proofs} provides the
corresponding analytical error bound $\Delta_{h_2}$ for the
mean-centered variance kernel $h_2$.

\subsection{Combinatorial Throughput Across Location
  (\texorpdfstring{$h_1$}{h1}), Scale (\texorpdfstring{$h_2$}{h2}), and
\texorpdfstring{$n \ge 64$}{n >= 64}}

Table~\ref{tab:benchmarks} reports the execution times and speedups
of both exact algorithms across the full benchmark suite for both
canonical kernels $h_1$ (location) and $h_2$ (sample variance) of
\citet{Serfling2002} and \citet{WildeGrimshaw2013}.

\begin{table}[t]
  \centering
  \caption{Empirical single-core execution time and speedup comparing
    Gosper's hack ($\Theta(k)$ per subset) and Knuth's Revolving Door
    Gray code ($\mathcal{O}(1)$ per subset) across both canonical
  kernels: Location ($h_1$) and Sample Variance ($h_2$).}
  \label{tab:benchmarks}
  \small
  \setlength{\tabcolsep}{3.8pt}
  \begin{tabular}{rrr|rrr|rrr}
    \toprule
    & & & \multicolumn{3}{c|}{\textbf{Location Kernel $h_1$ (Mean)}}
    & \multicolumn{3}{c}{\textbf{Scale Kernel $h_2$ (Sample Variance)}} \\
    $n$ & $k$ & Subsets $\binom{n}{k}$ & Gosper & RevDoor & Speedup
    ($h_1$) & Gosper & RevDoor & Speedup ($h_2$) \\
    \midrule
    20 & 3 &     1,140 &  0.01~ms & \textbf{0.01~ms} & $1.57\times$ &
    0.01~ms & \textbf{0.01~ms} & $1.03\times$ \\
    25 & 3 &     2,300 &  0.02~ms & \textbf{0.01~ms} & $1.36\times$ &
    0.01~ms & \textbf{0.01~ms} & $1.05\times$ \\
    29 & 5 &   118,755 & \textbf{0.74~ms} &  0.81~ms & $0.91\times$ &
    0.79~ms & \textbf{0.71~ms} & $\mathbf{1.12\times}$ \\
    35 & 5 &   324,632 &  2.23~ms & \textbf{2.01~ms} & $1.11\times$ &
    2.43~ms & \textbf{1.80~ms} & $\mathbf{1.35\times}$ \\
    40 & 5 &   658,008 &  3.42~ms & \textbf{3.28~ms} & $1.04\times$ &
    4.71~ms & \textbf{2.59~ms} & $\mathbf{1.82\times}$ \\
    50 & 5 & 2,118,760 & \textbf{11.07~ms} & 14.71~ms & $0.75\times$
    & 13.57~ms & \textbf{11.32~ms} & $\mathbf{1.20\times}$ \\
    25 & 7 &   480,700 &  2.86~ms & \textbf{1.75~ms} &
    $\mathbf{1.63\times}$ &  3.36~ms & \textbf{2.23~ms} &
    $\mathbf{1.51\times}$ \\
    30 & 7 & 2,035,800 & \textbf{10.25~ms} & 10.89~ms & $0.94\times$
    & 15.27~ms & \textbf{13.65~ms} & $\mathbf{1.12\times}$ \\
    \midrule
    80 & 3 &    82,160 & \texttt{---} & \textbf{0.49~ms} & \textit{$n
    \ge 64$} & \texttt{---} & \textbf{0.43~ms} & \textit{$n \ge
    64$} \\
    100 & 3 &   161,700 & \texttt{---} & \textbf{0.95~ms} &
    \textit{$n \ge 64$} & \texttt{---} & \textbf{0.79~ms} &
    \textit{$n \ge 64$} \\
    \bottomrule
  \end{tabular}
\end{table}

The empirical results in Table~\ref{tab:benchmarks} support the
structural comparison and default dispatch heuristic in
Table~\ref{tab:comparison}:
\begin{enumerate}
  \item \textbf{Sub-Millisecond Exact Resolution of Both
    \citet{WildeGrimshaw2013} Kernels ($h_1$ and $h_2$):} On the
    canonical $n = 29, k = 5$ benchmark ($118,755$ subsets),
    exhaustive evaluation completes in $0.74$~ms for the location
    kernel $h_1$ (via Gosper) and $0.71$~ms for the sample variance
    kernel $h_2$ (via Revolving Door). By comparison: (i)~exhaustive
    evaluation in Base~R via \texttt{median(utils::combn(x, 5, FUN =
    mean))} and \texttt{median(utils::combn(x, 5, FUN = var))}
    requires $244.2$~ms ($330\times$ slower) and $361.4$~ms
    ($509\times$ slower), respectively, while even the vectorized
    idiom \texttt{median(colMeans(utils::combn(x, 5)))} for $h_1$
    requires $44.5$~ms ($60\times$ slower) and allocates the full $5
    \times 118,755$ subset matrix in $\Theta(k M)$ memory;
    (ii)~unranking only the median $N^* = 32,160$ subsets for $h_1$
    and $N^* = 45,420$ subsets for $h_2$ via the custom R
    implementation of \texttt{kSubsetLexUnrank}
    \citep{KreherStinson1999} proposed by \citet[Table~3,
    p.~315]{WildeGrimshaw2013} at $\epsilon = 0.01$ takes $185.0$~ms
    ($250\times$ slower) and $261.2$~ms ($368\times$ slower) in R,
    and $1.50$~ms ($h_1$) and $2.20$~ms ($h_2$) in our direct
    \texttt{-O3} compiled-C unranking sweep
    (Section~\ref{sec:unranking}) even with a precomputed Pascal
    table and excluding sample-rank generation.
  \item \textbf{Consistent Advantage of Revolving Door on
    Second-Moment Kernels ($h_2$):} For $h_1$ at $k = 5$, Gosper's
    branch-free register loop ($5$ additions per subset) and
    Revolving Door ($2$ additions plus index branching) perform
    comparably. However, for the variance kernel $h_2$
    (Remark~\ref{rem:h2_kernel}), Gosper accumulates $2k = 10$
    additions across extracted bits per subset, whereas Revolving
    Door updates both shifted power sums $\Sigma_t^{(1)}$ and
    $\Sigma_t^{(2)}$ in $4$ arithmetic operations between
    recalibrations. Consequently, Revolving Door outperforms Gosper
    across all tested $(n, k)$ configurations for $h_2$ by
    $1.03\times$ to $1.82\times$, as well as by $1.63\times$
    ($274.1$~million subsets/sec) for $h_1$ at $n = 25, k = 7$.
  \item \textbf{Scaling Beyond the 64-Bit Word Limit ($n \ge 64$):}
    For $n = 80, k = 3$ ($M = 82,160$) and $n = 100, k = 3$ ($M =
      161,700$---matching the sample size $n = 100$ and kernel order $k
      = 3$ evaluated via the R algorithm of \citet{WildeGrimshaw2013}
    in \citet[Section~3]{FischerFriedWendler2016}), where single-word
    64-bit Gosper enumeration is inapplicable ($2^n \ge 2^{64}$),
    Revolving Door executes deterministically in $0.49$~ms /
    $0.43$~ms ($n = 80$) and $0.95$~ms / $0.79$~ms ($n = 100$) for
    $h_1$ and $h_2$, respectively.
\end{enumerate}

\subsection{Direct Unranking Crossover and Subsampling Error}
\label{sec:unranking}

\begin{figure}[t]
  \centering
  \includegraphics[width=\linewidth]{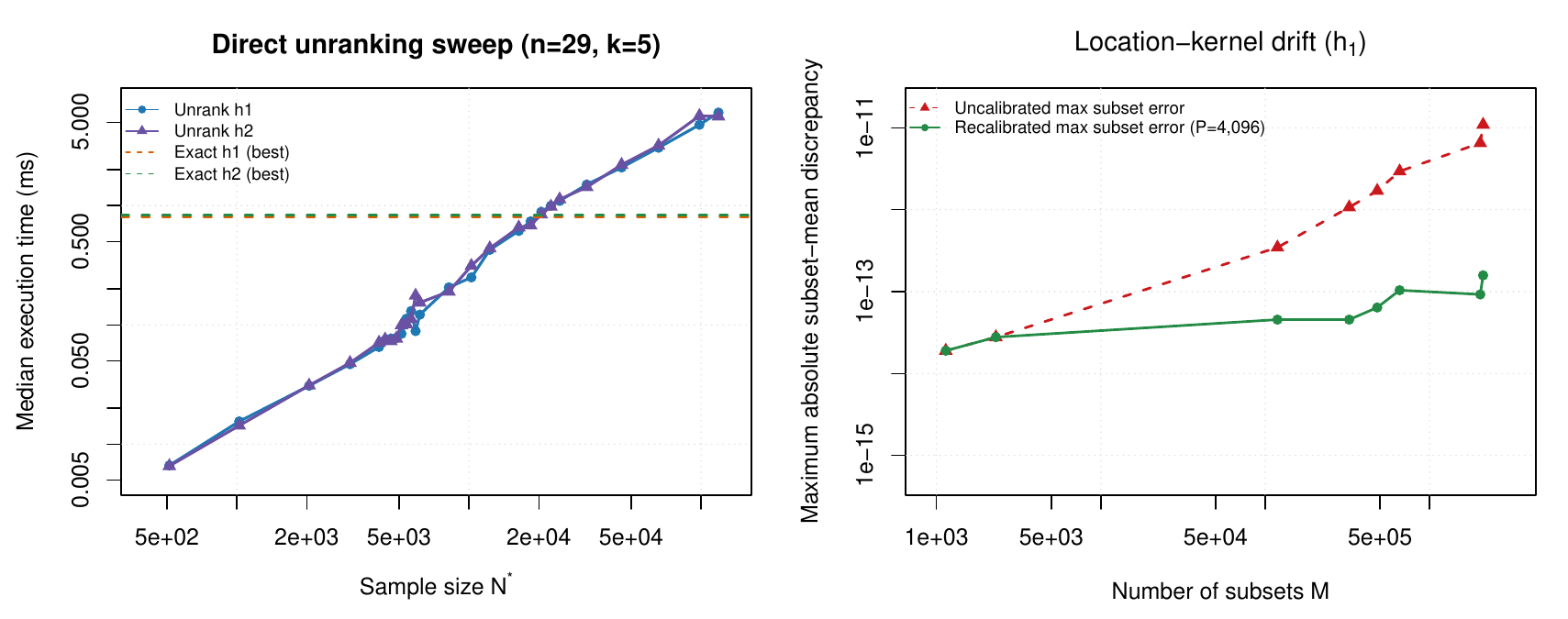}
  \caption{(a)~Direct compiled-C unranking sweep
    (\texttt{kSubsetLexUnrank}; precomputed Pascal table; selection
    included, rank generation excluded) versus exhaustive full
    enumeration of all $M = \binom{29}{5} = 118,755$ subsets on
    $n = 29, k = 5$, crossing over at $N^* \approx
    18{,}000$--$23{,}000$ subsets for both $h_1$ and $h_2$.
    (b)~Maximum observed location-kernel subset-mean discrepancy with
    and without periodic recalibration
    (Appendix~\ref{app:numerical-proofs}) across $M \in [1.14 \times
  10^3, 2.12 \times 10^6]$ ($P = 4,096$).}
  \label{fig:pareto}
\end{figure}

Figure~\ref{fig:pareto}(a) compares lexicographic subset unranking
(\texttt{kSubsetLexUnrank}) against exhaustive full-population
enumeration across the subsampling budget $N^* \in [512, 118{,}755]$
for $n = 29, k = 5$. To give unranking the most favorable compiled
baseline, the direct C sweep precomputes all binomial coefficients in
a Pascal lookup table and excludes random rank-sampling overhead from
the timer, measuring only subset unranking, kernel evaluation, and
order-statistic selection on the deterministic fixture $x_i =
10\sin(1.5i) + 0.2i$ ($i = 0, \dots, 28$). Even under these favorable
conditions in \texttt{-O3} compiled C, the cost of unranking $N^*$
subsets crosses exhaustive enumeration of all $M = 118,755$ subsets
at $N^* \approx 18{,}000$--$23{,}000$ subsets ($15\%$--$19\%$ of the
population) for both $h_1$ and $h_2$---well below the $N^* = 32,160$
($27.1\%$) and $N^* = 45,420$ ($38.2\%$) subsamples required by the
sequential stopping rule of \citet[Table~3,
p.~315]{WildeGrimshaw2013} at $\epsilon = 0.01$.

In $R = 500$ paired replications from the iid mixture
$0.9\,\mathcal{N}(0,1) + 0.1\,\mathcal{N}(5,0.5^2)$ at $n = 29, k =
5$, we compare squared errors relative to the uncontaminated location
$0$. The approximation draws observations without replacement within
each subset, but draws subsets with replacement across iterations;
its independently varied seeds are separate from the data-generation
stream. The empirical MSE is $0.291709$ for exhaustive
$\text{GM}_{29,5}$ and $0.297972$ for $B=500$ draws, a relative
difference of $2.15\%$ (approximate paired $95\%$ confidence interval
$0.45\%$--$3.85\%$, using the delta method). For $B=5,000$, the
empirical MSE is $0.293407$, a relative difference of $0.58\%$
(interval $0.07\%$--$1.09\%$). These intervals are pointwise
approximations; the target $0$ is the clean-component location, not
the mixture mean or median. In this separate R-package experiment,
single-fixture average CPU times were $4.008$~ms for exact
enumeration and $0.0494$~ms and $0.5655$~ms for $B=500$ and
$B=5,000$, respectively. These implementation-specific timings show
that smaller random samples can be faster, with an accuracy trade-off
distinct from the compiled-C unranking crossover above.

\subsection{Incremental Sliding-Window Evaluation}
\label{sec:sliding}

Unlike the preceding fixed-sample experiments, a sliding window
changes the sample after each new observation. Consecutive windows
share $n-1$ observations, so most subset kernels remain unchanged.
We maintain the $M=\binom{n}{k}$ kernel values with their ring-buffer
position masks. When the oldest observation in position $p$ is
replaced, only the $D=\binom{n-1}{k-1}$ values whose masks contain
$p$ need recomputation; the other $\binom{n-1}{k}$ values remain
unchanged. Selecting the upper median from a copy of all $M$ values
after each update then yields the exhaustive result, including for
duplicate observations and values. This extends the reuse principle
of online robust scale estimation
\citep{SchettlingerNunkesserFried2008,NunkesserFriedGather2009} to
generalized median kernels.

The prototype scans all masks and computes each affected kernel in
$\Theta(k)$ time: an update, including worst-case linear selection,
costs $\mathcal{O}(M+kD)$ time and $\mathcal{O}(M+n)$ space after
$\mathcal{O}(kM)$ initialization. Single-word masks limit this
implementation to $n\le63$. For $h_2$, a two-pass variance
computation around a subset observation avoids accumulated
cross-window recurrence drift, although floating-point rounding remains.

Tests compare each update with direct combination enumeration and
full recomputation, covering ties, even populations, large offsets,
boundary cases and 5,000-update sequences. For the benchmarks, every
update was also checked against both Gosper and Revolving Door. On
Apple M3 Pro, Apple Clang~21.0.0.1.1 (\texttt{-O3}), the stream
$x_t=10\sin(1.5t)+0.02t$ was measured in two runs of seven 64-update
batches. Sliding-state initialization was excluded; full
recomputation included its per-call allocations. For $n=25,k=7$,
median batch times per update were $2.994$--$3.032$~ms (sliding
$h_1$) versus $3.556$--$3.599$~ms (full Revolving Door), and
$3.258$--$3.276$ versus $3.528$--$3.552$~ms for $h_2$. For
$n=29,k=5$, sliding was not consistently faster. The scan and
repeated selection still cost $\mathcal{O}(M)$ per update.

\section{Software Availability}
\label{sec:software}

The self-contained ANSI C reference implementation implementing both
Gosper's bit-parallel enumerator and Knuth's Revolving Door Gray code
with periodic drift recalibration for the location ($h_1$) and scale
($h_2$) kernels, together with all benchmark reproduction scripts
used in this study, is available from the author upon reasonable
request. Because no dedicated package for generalized median
estimators ($\text{GM}_{n,k}^{(h)}$ with $k \ge 3$) currently exists
on CRAN or PyPI, open-source software packages for R and Python
featuring the automatic $(n, k, h)$ dispatch rule established in
Table~\ref{tab:comparison} are currently in preparation and will be
released publicly alongside the extended journal version of this work.

\section{Conclusions and Outlook}
\label{sec:conclusion}

We have established both theoretically and empirically that
generalized median estimators ($\text{GM}_{n,k}^{(h)}$) for location
($h_1$) and scale ($h_2$) can be evaluated exhaustively in
$\mathcal{O}(M + kM/P)$ worst-case time (plus $\mathcal{O}(k)$ setup)
and $\mathcal{O}(k + M)$ space, refuting the long-standing assumption
in computational statistics that subset counts on the order of $10^5$
to $10^6$ require stochastic subsampling. From an algorithmic
perspective, replacing the expected $\Theta(n)$-per-subset scan of
uniformly sampled lexicographic unranking with stateful Word-RAM
traversal ($\Theta(k)$ per subset via Gosper's recurrence for $n <
w$) or Revolving Door Gray codes ($\mathcal{O}(1)$ per subset between
  recalibrations via single-element exchanges $|S_t \triangle S_{t+1}|
= 2$ for arbitrary $n \in \mathbb{N}$) eliminates the combinatorial
unranking bottleneck, while periodic anchor recalibration bounds
floating-point rounding error for both $h_1$ and $h_2$, independently
of $M = \binom{n}{k}$, under the assumptions stated in
Appendix~\ref{app:numerical-proofs}.
From an algorithm-engineering perspective, our structural comparison
shows that Gosper and Revolving Door are complementary: Gosper excels
in the branch-free single-register regime ($n \le 63, k \le 5, h =
h_1$), whereas Revolving Door is advantageous for $h_2$ in the tested
cases and supports sample sizes $n \ge 64$; for larger $k$,
performance depends on the instance, resolving the classic $n = 29, k
= 5$ benchmark of \citet{WildeGrimshaw2013} in $0.74$~ms ($h_1$) and
$0.71$~ms ($h_2$). For sliding windows, Section~\ref{sec:sliding}
reuses unchanged kernel values and updates only the affected subsets.
Future work will investigate dynamic order-statistic structures for
avoiding full reselection, together with coarse-grained multi-core
parallelization by invoking revolving door unranking once per thread
to partition $\mathcal{S}_{n,k}$ into $T$ contiguous Gray code segments.

\appendix
\section{Numerical Error Bounds and Proofs}
\label{app:numerical-proofs}

\subsection{Location-Kernel Error Bound}

\begin{proposition}[Location-Kernel Error under Periodic Recalibration]
  \label{prop:drift}
  Let $X_{\max} = \max_{0 \le i < n} |x_i|$, let $P \in
  \mathbb{N}_{\ge 1}$ denote the recalibration period in
  Algorithm~\ref{alg:revdoor}, and set $L = k + 2(P - 1)$. In the
  standard floating-point model with unit roundoff $u =
  \varepsilon_{\text{mach}}$ ($Lu < 1$, with no overflow or underflow):
  \begin{enumerate}
    \item For every subset index $t \in \{0, \dots, M-1\}$, the
      number of incremental transitions since the most recent direct
      summation is $r_t = t \bmod P \le P - 1$, independent of $M =
      \binom{n}{k}$.
    \item For the location kernel $h_1$, the computed subset mean
      $\widehat{\mu}_t = \operatorname{fl}(\widehat{\Sigma}_t / k)$
      and the exact real subset mean $\mu_t = \frac{1}{k}\sum_{j \in
      S_t} x_j$ satisfy:
      \begin{equation}
        \label{eq:drift_bound}
        \left| \widehat{\mu}_t - \mu_t \right| \le \left( k + 2(P -
        1) \right) \varepsilon_{\text{mach}} X_{\max} +
        \mathcal{O}(\varepsilon_{\text{mach}}^2),
      \end{equation}
      at an amortized recalibration overhead of $k / P$ additions per subset.
  \end{enumerate}
\end{proposition}

\begin{proof}
  At every anchor step $t_0 = q P$ ($q \in \mathbb{N}_0$),
  $\widehat{\Sigma}_{t_0}$ is recomputed directly by summing $k$
  terms of magnitude at most $X_{\max}$, incurring standard summation
  error $|\widehat{\Sigma}_{t_0} - \Sigma_{t_0}| \le (k -
  1)\varepsilon_{\text{mach}} k X_{\max} +
  \mathcal{O}(\varepsilon_{\text{mach}}^2)$. Each of the $r_t \le P -
  1$ subsequent incremental updates subtracts $x_{u_{t_0+s}}$ and
  adds $x_{v_{t_0+s}}$ while $|\Sigma_{t_0+s}| \le k X_{\max}$,
  contributing at most $2 k \varepsilon_{\text{mach}} X_{\max}$ error
  per step. Dividing by $k$ (adding at most one final relative
  rounding error $\varepsilon_{\text{mach}} X_{\max}$) yields
  Eq.~\eqref{eq:drift_bound}, while direct recomputation performs $k$
  additions once every $P$ subsets ($k / P$ additions amortized per subset).
\end{proof}

\subsection{Scale-Kernel Error Bound}

\begin{proposition}[Scale-Kernel Error under Periodic Recalibration]
  \label{prop:h2_error}
  Assume finite binary floating-point inputs, the standard model
  $\operatorname{fl}(a\circ b)=(a\circ b)(1+\delta)$ with
  $|\delta|\le u=\varepsilon_{\text{mach}}$, no overflow or
  underflow, and no unsafe reassociation of the operations in
  Eq.~\eqref{eq:h2_update}. Let $c$ be the computed reference mean,
  $a_i=x_i-c$, $d_i=\operatorname{fl}(a_i)$, and
  $q_i=\operatorname{fl}(d_i^2)$, and define $A=\max_i|a_i|$,
  $D=\max_i|d_i|$, and $Q=\max_i|q_i|$. Let $P$ be the recalibration
  period, $L=k+2(P-1)$, $\gamma_m=mu/(1-mu)$, and assume $Lu<1$. Set
  \begin{equation}
    \label{eq:h2_error_terms}
    E_1=\gamma_{L-1}LD,\quad E_2=\gamma_{L-1}LQ,\quad
    B_1=kD+E_1,\quad B_2=kQ+E_2,\quad
    C=B_1^2/k,\quad E_p=\gamma_3C,\quad E_s=u(B_2+C+E_p).
  \end{equation}
  For every subset $S$ of size $k$, the computed variance-kernel
  value $\widehat h_2(S)$ satisfies
  \begin{equation}
    \label{eq:h2_error_bound}
    \left|\widehat h_2(S)-h_2(S)\right| \le
    \frac{(2u+u^2)kA^2+ukD^2+E_2+E_1(2kD+E_1)/k}{k-1}
    +\frac{E_p+E_s+\gamma_2(B_2+C+E_p+E_s)}{k-1}
    \equiv \Delta_{h_2}.
  \end{equation}
  Consequently, the computed upper median differs from the exact
  real-arithmetic upper median by at most $\Delta_{h_2}$.
\end{proposition}

\begin{proof}
  Variance is invariant under a common translation, so the exact
  kernel is unchanged by replacing each $x_i$ by $a_i=x_i-c$,
  irrespective of the rounding error in the computed reference mean
  $c$. The centering operation gives $d_i=a_i+e_i$ with $|e_i|\le
  u|a_i|$. Writing the sample variance as $\|H a_S\|_2^2/(k-1)$,
  where $H$ is the orthogonal projection that removes the subset
  mean, and using $\|H\|_2\le1$, we obtain
  $|\|H d_S\|_2^2-\|H a_S\|_2^2|\le(2u+u^2)\|a_S\|_2^2\le(2u+u^2)kA^2$.
  Rounding the individual squares contributes at most $ukD^2$ to the numerator.

  Between anchors, each running sum is formed from at most
  $k+2(P-1)=L$ signed terms. The standard sequential-summation bound
  therefore gives absolute errors at most $E_1$ and $E_2$ for the
  sums of $d_i$ and $q_i$, respectively. The resulting errors in the
  unrounded numerator are bounded by $E_2+E_1(2kD+E_1)/k$. Finally,
  the rounded square, reciprocal multiplication, subtraction, and
  final reciprocal multiplication in Eq.~\eqref{eq:h2_update}
  contribute at most the second term of
  Eq.~\eqref{eq:h2_error_bound}, by the same floating-point model.
  Dividing by $k-1$ gives the per-subset bound. An order statistic is
  1-Lipschitz in the maximum norm of its input vector; applying this
  fact to the $M$ per-subset bounds proves the median statement.
\end{proof}

\bibliographystyle{elsarticle-harv}
\bibliography{ExactGeneralizedMedian}

\end{document}